\documentclass[]{spie}  

\usepackage{amsmath,amsfonts,amssymb}
\usepackage{graphicx}
\usepackage[colorlinks=true, allcolors=blue]{hyperref}

\title{The optimisation of short-term scheduling of science observations at Paranal observatory (VLT and ELT)}

\author[a]{Joseph P. Anderson}
\author[a]{Elyar Sedaghati}
\author[b]{Aleksandar Cikota}
\author[a]{Natalie Behara}
\author[a]{Fuyan Bian}
\author[a]{Angel Otarola}
\author[a]{Steffen Mieske}
\affil[a]{European Southern Observatory, Alonso de C\'ordova 3107, Casilla 19, Santiago, Chile}
\affil[b]{Gemini Observatory/NSF’s NOIRLab, Casilla 603, La Serena, Chile}

\authorinfo{Further author information: (Send correspondence to J.P.A)\\J.P.A. E-mail: janderso@eso.org}

\begin{document} 
\maketitle

\begin{abstract}
The efficiency of science observation Short-Term Scheduling (STS) can be defined as being a function of how many highly ranked observations are completed per unit time. Current STS at ESO's Paranal observatory is achieved through filtering and ranking observations via well-defined algorithms, leading to a proposed observation at time t. This Paranal STS model has been successfully employed for more than a decade.
Here, we summarise the current VLT(I) STS model, and outline ongoing efforts of optimising the scientific return of both the VLT(I) and future ELT. We describe the STS simulator we have built that enables us to evaluate how changes in model assumptions affect STS effectiveness. Such changes include: using short-term predictions of atmospheric parameters instead of assuming their constant time evolution; assessing how the ranking weights on different observation parameters can be changed to optimise the scheduling; changing STS to be more `dynamic' to consider medium-term scheduling constraints. We present specific results comparing how machine learning predictions of the seeing can improve STS efficiency when compared to the current model of using the last 10\,min median of the measured seeing for observation selection. 
\end{abstract}


\section{INTRODUCTION}
\label{sec:intro}

The La Silla Paranal Observatory's (LPO's) Paranal Science Operations (PSO) uses its Observing Tool (OT) to filter, rank and then select the optimal Observing Block (OB) from a database of all valid Service Mode, SM, OBs for the current semester, to execute at each telescope at any given moment. (Note, this ignores visitor mode VM, and Target of Opportunity, TOO, runs.)  The ranking uses as input: a) deterministic time-related constraints from the OB: coordinates, airmass limit, Moon FLI, time constraints; b) atmospheric constraints from the OB: seeing, coherence time, ground layer faction (GLF), sky transparency, precipitable water vapour (PWV); c) restrictions on telescope pointing from the wind; d) scientific ranking; e) accumulative probability distributions for the constraints (both deterministic, e.g., moon FLI, and empirical, e.g., seeing); and f) a defined rule to combine scientific ranking and probability that observing constraints are met. The ranking algorithm then provides a ranked list of OBs to be executed at the given input time, with the top ranked OB being selected for execution (after a manual check), provided that it requires the same instrument that is currently at telescope focus. We refer to this process of OB selection and execution as ‘Short Term Scheduling’; STS. (See Ref.\,\cite{sil01} for an early description of ESO's SM scheduling and execution, and Ref.\,\cite{bie10} for more details on the filtering and ranking methodology\footnote{Additional information on the general philosophy of ESO's SM observing can be found on the public, \href{https://www.eso.org/sci/observing/phase2/SMPhilosophy.html}{science-users website.}}.)

Under the current PSO STS operations model the measured atmospheric parameters are automatically ingested, and a median of the last 10\,min is used as input for the filtering of OBs. An OB is thus selected based on these current conditions. We refer to this current STS model as ‘precast’. After its execution the quality control of the data (QC0) is performed, depending mostly on the evolution of the atmospheric parameters (or in some cases their influence on the properties of the data products) during the science exposures. If the OB passes QC0 then it is graded ‘A’ if all constraints are met, or ‘B’ if constraints are almost met, and the execution is considered a success. If an OB fails at QC0 it is graded ‘C’ (constraints are not met) and it is returned to the database of observations to be attempted again - it obtains an `M', `must repeat' status. 

In the future (e.g. as a requirement of ESO's ELT) it is envisaged that the atmospheric input to the ranking and filtering of observations will be supported by forecast of the atmospheric conditions (provided from, e.g. machine-learning nowcasting, numerical weather prediction models, or a combination of both). Several efforts are currently ongoing at ESO to provide such forecasts, and we describe and test one such effort - seeing nowcasting - in this manuscript. 
The usefulness of the integration of any forecast into the STS model depends on at least two parameters having different uses for interpreting the gain produced by their integration: 1) the predictive power of the forecast compared to current precast; 2) the improvement of STS efficiency when using the forecast. The former is independent of the observing constraints of OBs in the queue, or indeed the current state of the queue (e.g. the distribution of OBs with different properties). The latter strongly depends on the properties of the OBs available to execute (this will become clear when we present our results below, specifically in Section~\ref{empiricalres}).
In the current contribution we assess whether machine-learning (ML) predictions of the seeing (nowcasting) can improve STS efficiency at Paranal observatory. First, in the next section we present `observation execution statistics' that document how efficient the current PSO STS model is at successfully executing science OBs, with respect to changing atmospheric conditions. Then, in Section\,\ref{secsimulator} we describe our STS simulator, built to quantitatively assess how changes in the STS model affect STS efficiency. In Section\,\ref{secnowcast} we briefly present the seeing nowcasting model to be tested. The simulations to analyse different PSO STS models are outlined in Section~\ref{secstssim}. These models are then tested - using the simulator - and the results are presented in Section~\ref{results}. Finally in Section~\ref{conclusions} we provide our conclusions and an outlook for how our simulator will be used over the coming years to further optimise science OB execution.

\section{Observation execution statistics}
\label{stats}
A major motivation for producing seeing predictions, and then testing them in the current analysis, is that a small but non-negligible number of observations are ‘lost’ because the atmospheric conditions evolve to be out of constraints. In this section we thus analyse execution statistics for all VLT instruments between periods P97 and P103 (April 2016 to September 2019). Specifically, we concentrate on the ‘M’ must repeat fraction: the number (and percentage) of hrs ‘lost’ because an OB was started but then graded C at QC0 and thus returned to the observing queues for later observation. Whenever this happens due to an evolution in atmospheric parameters, it can be considered a failure of the PSO STS precast model (together with the current \href{https://www.eso.org/sci/facilities/paranal/quality-control/qc0-ob-grading.html}{PSO QC0 policy}), and a loss of telescope time. Below, we proceed to document the absolute and relative amount of ‘lost’ telescope time as a function of VLT instrument. 

\begin{figure} [ht]
\begin{center}
\begin{tabular}{c}
\includegraphics[height=8cm]{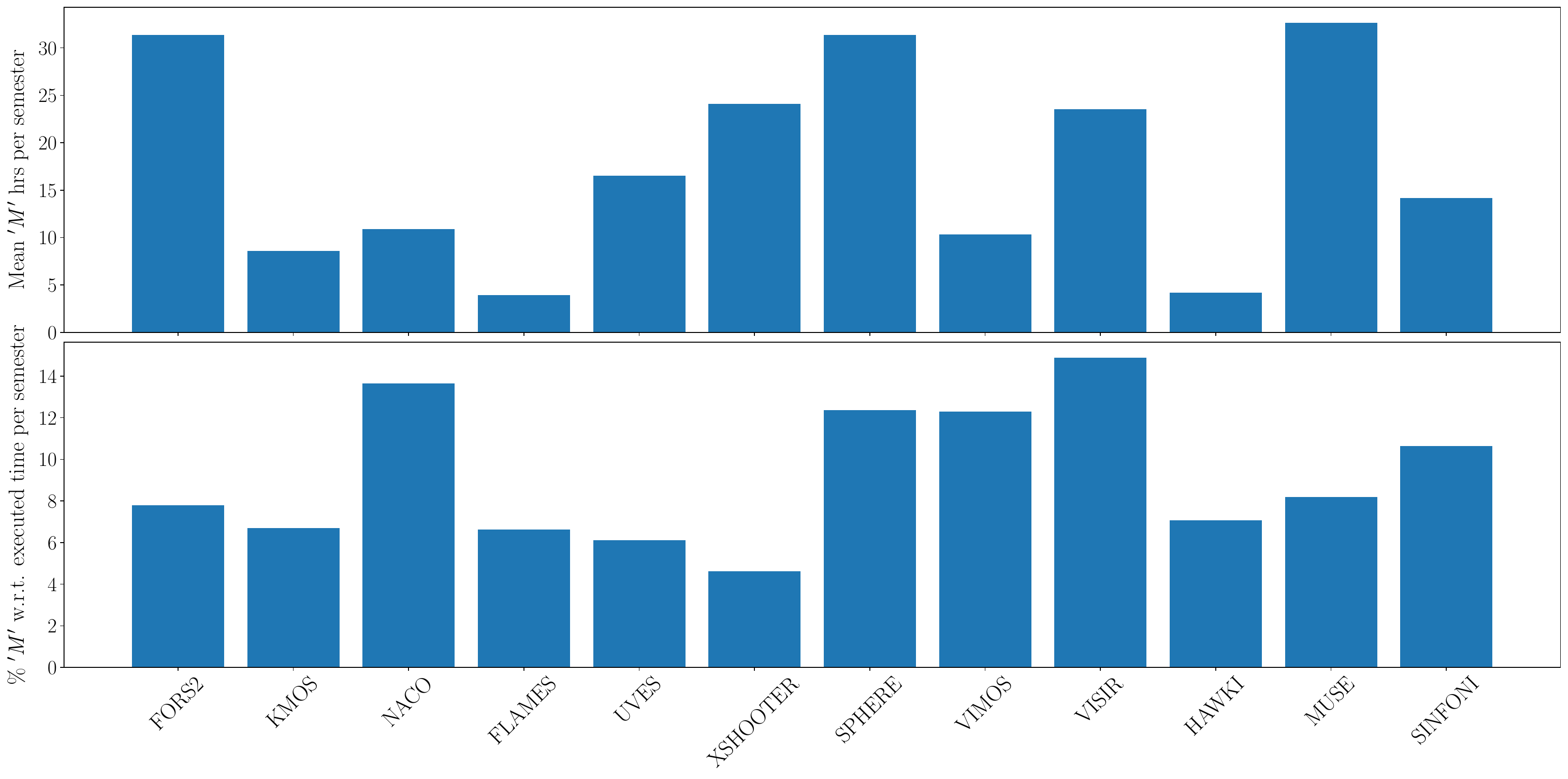}
\end{tabular}
\end{center}
\caption
{\label{Mhrs} 
\textit{(Top)} Average number of hours of telescope time lost per semester on each VLT instrument. (\textit{Bottom}) Average number of hours of telescope time lost per semester on each VLT instrument as a percentage of the total execution time at the instrument. Instrument names are labeled on the x-axis}
\end{figure}


First, we analyse M as a function of total hours lost on the different VLT instruments per observing semester (each of six months in duration). Figure\,\ref{Mhrs} shows the M repeat hours per semester per instrument. The most significant offenders are: FORS2\cite{app98}\,; XShooter\cite{ver11}\,; SPHERE\cite{beu19}\,; VISIR\cite{lag04}\,; and MUSE\cite{bac10}. However, some instruments have many more allocated runs and thus many more attempted OB executions than others. Therefore, Figure\,\ref{Mhrs} also shows the hours lost per instrument as a percentage of the total execution time of science observations on that instrument per semester. This metric informs us how problematic each instrument is in terms of OBs being graded C - independent of the total hrs of possible executions. The overall M \%, averaged over all UT instruments is around 10\%. Figure~\ref{Mhrs} shows that the ‘workhorse’ instruments FORS2, XShooter, and MUSE are now relatively low offenders with respect to the total time of OBs that are in their queues. They have an M \% of 7.8, 4.6 and 8.2 respectively. The worst instruments after this normalisation are NACO\cite{len03} (now decommissioned), SPHERE, VISIR and SINFONI\cite{eis03} (decommissioned) together with VIMOS\cite{lef03} (decommissioned). NACO, SPHERE and SINFONI are adaptive optics (AO) instruments while VISIR works in the mid-infrared. If we separate instruments into AO and non-AO (removing VISIR due to its distinct operating wavelength), we find that the AO average is 12.2\%, while the non-AO average is 8.7\%. There is a clear difference in that AO instruments have a significantly higher M fraction than non-AO instruments. This difference may have important implications for OB executions at the VLT UT4 (now an ‘AO telescope’) and future ELT instruments (where the telescope and all instruments will be of AO nature).

While there are several atmospheric parameters that could evolve to be out of constraints - seeing, sky transparency, precipitable water vapour, etc, observing experience suggests that changes in seeing are the dominant cause of an unsuccessful science observation, especially for non-AO instruments. To test this intuition, we analysed all OB executions during 2021 (downloaded from Paranal's Night-Log Tool). The total fraction of C-graded OBs due to seeing in 2021 was more than 50\%. This was an important finding for the work documented in this contribution, as it justifies our initial focus on seeing. 

In conclusion, around 10\% of telescope time is lost due to atmospheric conditions evolving out of constraints. Additionally – and of utmost importance for the analysis in the current document – more than half of this time is due to the seeing evolving out of constraints. This motivates the below analysis, where we test whether this amount of lost telescope time could be decreased by using nowcast seeing predictions in place of the precast model. 
\begin{figure} [ht]
\begin{center}
\begin{tabular}{c}
\includegraphics[height=11.5cm]{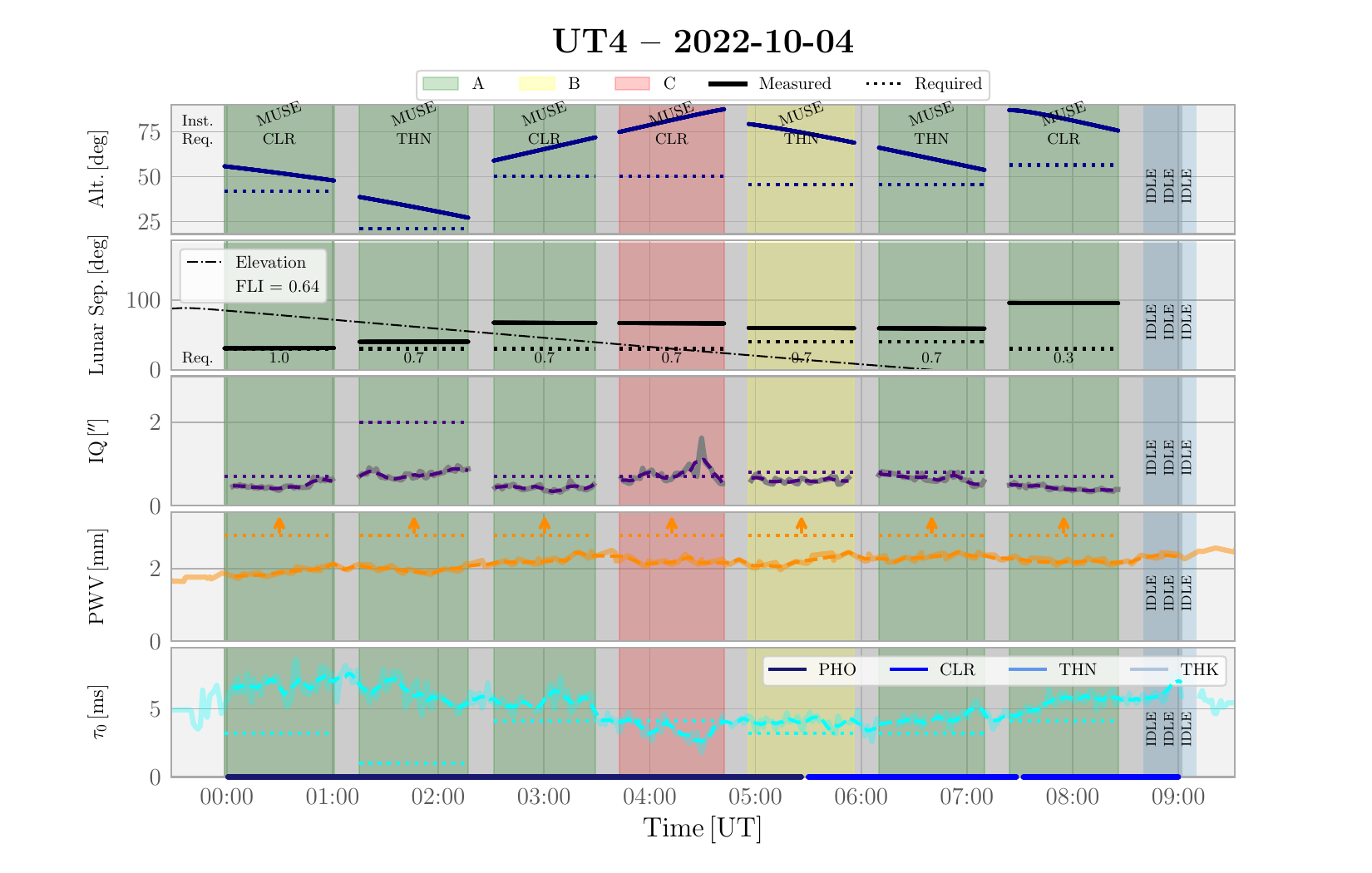}
\end{tabular}
\end{center}
\caption 
{\label{figsimulator} 
A visualisation of our STS simulator during one night at the UT4 telescope. The figure is separated into five rows, each showing the observing condition requirement (dashed horizontal lines) and the measured parameter (solid lines). The x-axis shows the time in UT. The top panel shows the altitude of the observation (i.e. the airmass). This panel also notes the instrument being used and the required sky transparency (with the measured transparency indicated by the narrow horizontal line at the bottom of the plot). The next panel down gives the lunar separation and the FLI of that night. The required FLI of each OB is given at the bottom of the panel. The middle panel gives the Image Quality (IQ) constraint and measured value (the latter being converted from the measured seeing). The second to bottom panel displays the precipitable water vapour (the constraints are worse than the measured value in this example and thus all constraints are shown with arrows indicating that they are higher than the upper limit of the panel y-axis). Finally, in the bottom panel the required and measured coherence time are plotted. The grey gaps between OBs reflect the time added to the simulation to include a combination of on-sky instrument calibration time together with various `losses' that occur during real-life night-time operations (these are added to ensure that the total time for science OB executions is realistic).}
\end{figure}

\section{A Paranal Short-Term Scheduling simulator}
\label{secsimulator}
The project driving the current manuscript started by analysing historical OB execution statistics, some of which was outlined in the previous section. However, it was quickly concluded that to robustly and quantitatively assess the effects of any changes in the STS model, an STS simulator was required. Simulating such a process allows one to separate individual model parameter changes, something which is much more difficult when analysing historical data where there is significant uncertainty on why a specific observation was successful or not, and whether the implemented observing strategy was optimal. 

Our STS simulator has been built to replicate PSO STS of OBs as robustly as possible. This means selecting an observation from a list of valid OBs for execution at time t, through filtering and ranking of OBs using parameters within the OB (coordinates, observing constraints) and the measured current atmospheric conditions. After selection of an OB, this observation is virtually executed, and QC0 is performed on its completion using the evolution of the relevant atmospheric parameters for the duration of the OB. A successful OB is removed from the OB queue, while an unsuccessful OB (where the atmospheric parameters evolved out of constraints) is returned to the queue for future execution.  

The simulator uses as input: real OB ‘queue dumps’ at the start of each observing semester (all approved observations) at each UT telescope; time series of atmospheric data; sky transparency conditions; and predictions of the seeing from the nowcast. This input is used together with probability distributions of different ranking parameters (e.g. seeing, time until object will set, FLI, etc), and a QC0 model that determines whether an OB execution is considered successful. The simulator then runs through a semester, selecting and ‘executing’ successive OBs at each time t that the previous OB finishes. At the end of a semester one can then assess the efficiency of the STS model. To test any aspect of the PSO STS model, changes in any given parameter can be made and the effect of that change can be evaluated. To do this, we define the baseline STS model being that where all parameters follow current practice at PSO: how OBs are filtered and ranked and how QC0 is applied. Figure~\ref{figsimulator} shows one simulated night of observations (04/10/2022), using the standard baseline model (precast) at the UT4 telescope, that hosted three instruments at that time: MUSE, HAWK-I\cite{pir04} and ERIS\cite{dav23} (however no ERIS OB was executed on that night). The plot visualises how the simulator starts at the end of evening twilight, selects a first OB for execution (in this case from MUSE), follows the observing conditions through the execution time of the OB, then uses the latter to evaluate the success of the OB when it finishes (QC0). The simulator then continues this process through the night until the start of morning twilight. At the end of one night, the simulator moves forward in time to the end of evening twilight of the subsequent night and repeats. The simulator runs over all nights of each semester and OB execution statistics are obtained and can be analysed. 

\begin{figure} [ht]
\begin{center}
\begin{tabular}{c}
\includegraphics[height=8cm]{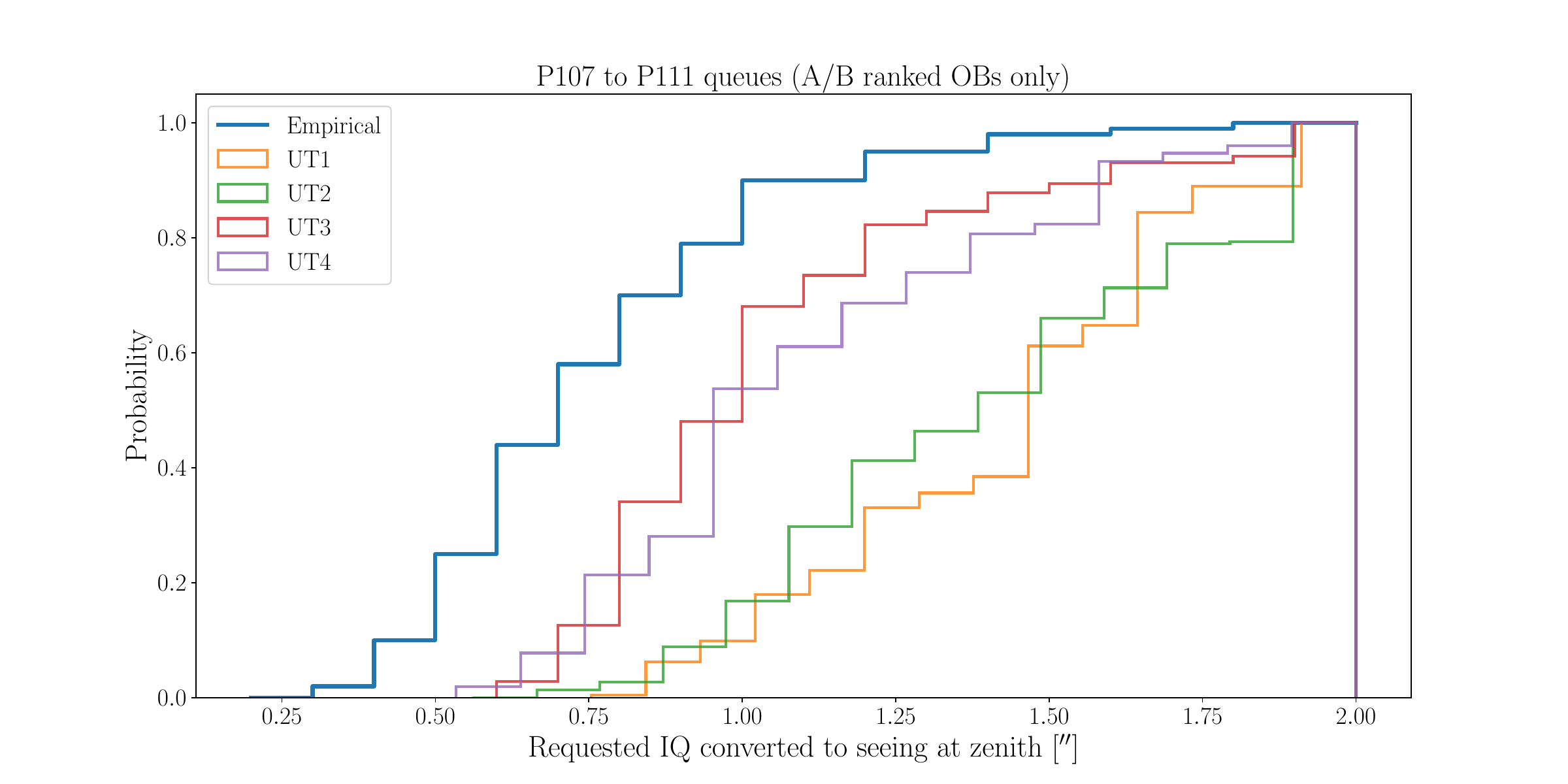}
\end{tabular}
\end{center}
\caption 
{\label{seeing}
Cumulative distributions of the measured `empirical' seeing at Paranal, together with the requested distributions in semesters between 2020 and 2022. The blue distribution gives that measured at Paranal observatory. This distribution has a median value of around 0.7 arcseconds. The other colours show the seeing constraints (converted from requested IQ) of all OBs for each semester (all UT telescopes combined). The cumulative distribution of the requested seeing constraints are significantly offset to higher seeing than that `available' during a semester, with a median offset of around 0.5 arcsecond.}
\end{figure}

Below we test the baseline PSO standard model – that applies precast seeing – against a model that uses a nowcast seeing prediction as input. We present two sets of simulations – both comparing the results of using precast vs nowcast seeing input. The first uses real OBs and their seeing constraints. The second replaces each OB seeing constraint with a seeing value randomly drawn from the empirical distribution of seeing as measured at Paranal. These second simulations are motivated by the observation that the user-requested seeing values are significantly offset to higher values than the empirical distribution, as shown in Figure~\ref{seeing}. The second simulation set is then used as a first order test of how STS will perform in the ELT era, where we assume that users will request conditions that more closely match those available. 
Before describing the simulations in more detail and analysing their results, the next section briefly summarises the nowcast seeing predictions.

\section{Nowcasting: machine-learning seeing predictions}
\label{secnowcast}
The main initial aim of our work was to investigate whether a forecast of the seeing could perform better than assuming the seeing stays the same during an OB execution (precast). 
Therefore, our first simulations presented below focus on testing ML learning predictions of the seeing; the nowcast. Seeing nowcasting has been under investigation for several years at ESO (see Ref.\,\cite{mil20} for previous work internal to ESO, and e.g.\, Ref.\,\cite{mas23} for other seeing forecast efforts). A full description of such modelling/methods will not be described here, as a separate detailed analysis will be presented in the future. 
The specific model used below uses the Random Forrest type of ML. In very basic terms, the nowcast uses as input, time series of: seeing, coherence time, ambient pressure and temperature, wind speed and direction, together with the time since sunset, 
as training sets between a range of dates to predict the seeing in the next hour of the night after time t when an OB is scheduled to be started. Data is separated by observing semester, with all predicted semesters only using data from the past as training sets. 

With seeing predictions from the nowcast, we can compare the efficiency of using such a seeing model in contrast to using the last 10\,min median precast model currently in use for PSO STS. Thus, within our simulator at every OB selection time t, we can select an OB following precast or nowcast. Simulating this across many OBs during different semesters, telescopes and instruments, allows us to robustly test the nowcast in an operationally realistic environment.

\section{STS simulations}
\label{secstssim}
\subsection{The baseline PSO STS model - precast, and nowcast}
\label{base}
As discussed above, our simulation methodology is to define our baseline model that is consistent with current PSO STS operations, then contrast any new model against our current standard practice, evaluating whether the new model performs better or worse than that currently employed. Using such methods we can test all the existing parameters in the PSO STS operations model and attempt to optimise their usage. Precast is our current model, where the last 10\,min median of measured atmospheric conditions are used to filter out OBs. Then, a ranking of remaining OBs is achieved using using defined probability distributions for observing conditions together with the scientific ranking of each OB (defined by the rank of the proposal run). QC0 is then evaluated using the average observing conditions throughout the duration of the science observation. While there are many parameters within this standard precast model that could be changed and tested, our current focus is to test the usage of nowcast seeing predictions in place of the precast seeing. Therefore, the nowcast simulations presented below maintain all other parameters the same (including using the last 10\,min median for input of all atmospheric parameters \textit{except} seeing), except the input seeing where nowcast predictions are employed.

Simulations are run at all four VLT UTs, together with their installed set of instruments, during all semesters between P107 to P111 (thus those OBs approved for execution at the start of each semester between April 2021 and September 2023).

\begin{figure} [ht]
\begin{center}
\begin{tabular}{c}
\includegraphics[height=10cm]{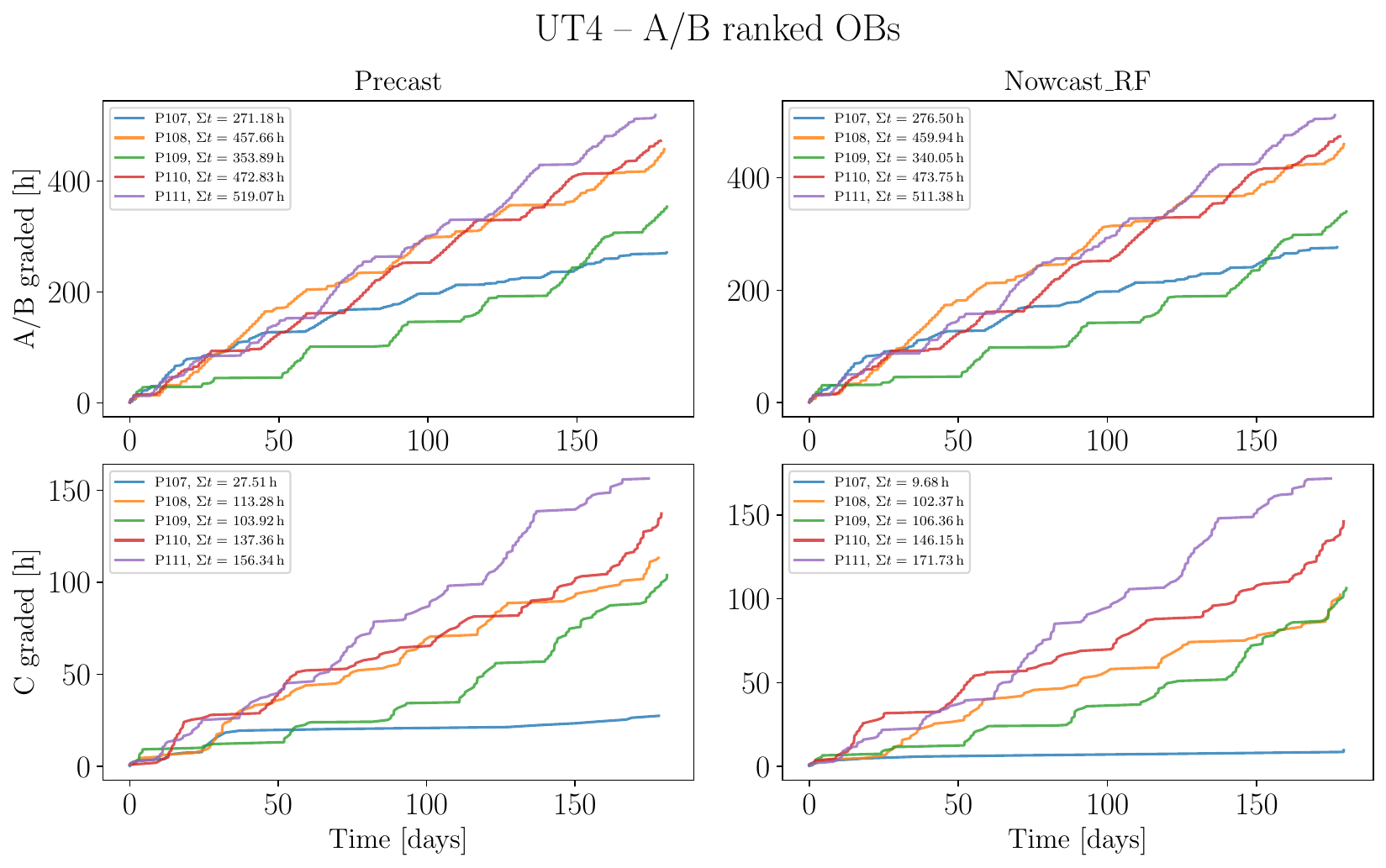}
\end{tabular}
\end{center}
\caption 
{\label{simulation} 
\label{metrics}
Cumulative distributions of simulated OB executions for the standard PSO precast model (left) and the nowcasting seeing model (right). The top panels show the cumulative distributions of successfully (A or B graded) A- and B-ranked OBs, while the bottom panels display the cumulative distributions of the unsuccessful (C graded) A- and B-ranked OBs. The x-axis on each plot is time in days, whle the y-axis is the number of hours of OBs executed. Large differences are observed in different semesters.}
\end{figure} 

\subsection{STS metrics}
\label{sec_metrics}
To assess the success of any given model, metrics must be defined. An overall high-level metric to measure the success of STS can be defined as: successfully executing as many highly ranked science OBs per unit time. This flows down from an ESO Council document\footnote{\href{https://www.eso.org/public/about-eso/committees/cou/cou-154th/external/Cou_1847_rev_Science_Policies_050520.pdf}{ESO Optica/Infrared Telescopes Science Operations Policies.}}that states: ``The scheduling process follows the scientific rank of the proposals and seeks to optimise the scientific return." In this context, it is relevant to state that approved observing runs are separated into three categories that depend on the scientific ranking of proposals. A rank are the highest scientifically ranked runs. ESO makes every effort to execute observations of A-ranked runs, and the possibility exists to carry such OBs over from one semester to the next, if observations are still pending. Next are B-rank runs. ESO again attempts to observe all observations for such runs during a semester, but if this is not possible then observations for B-ranked runs are not carried over to the next semester. The last are C-rank runs; those that have not been ranked sufficiently highly to be scheduled for execution as A or B rank, but that have sufficiently relaxed observing constraints that the OBs are scheduled for observation as `fillers' - observations that are only executed when there are no other A- or B-ranked OBs in the queue for execution at a specific time in sub-optimal observing conditions.

In the current work we defined two metrics: 1) the hours of successfully A- and B-ranked OBs that are completed each semester, and 2) the hours of C-graded (unsuccessful) A- and B-ranked OBs executed each semester (therefore, C-ranked observations are ignored for our initial metrics used in this work)\footnote{To avoid confusion, it is important to explicitly state that the nomenclature `A/B/C' is used twice. First, it is used to define the scientific ranking of runs and therefore individual observations; the `rank'. Second, it is used to give the QC0 of individual OB executions; the `grade'.}. The ‘best’ model should maximise 1) while minimising 2). 
The currently defined metrics are quite simple and do not include other important parameters such as: run completion, completion of OBs as a function of requested constraints, idle time when the STS model does not find any suitable OB for execution, amongst other parameters. Such more encompassing metrics will be included in subsequent work. 

Using the above defined metrics, we can then evaluate the relative success of precast vs nowcast over all UTs and all simulated semesters. The simulation results are depicted in Figure~\ref{simulation} for UT4 in terms of cumulative OB execution distributions. One can see that results are quite dependent on semester. Each semester the OBs available for execution are changing, as is the time evolution of the observing conditions.

\section{Results}
\label{results}
\subsection{Nowcast vs Precast}
\label{res1}

\begin{figure} [ht]
\begin{center}
\begin{tabular}{c}
\includegraphics[height=11cm]{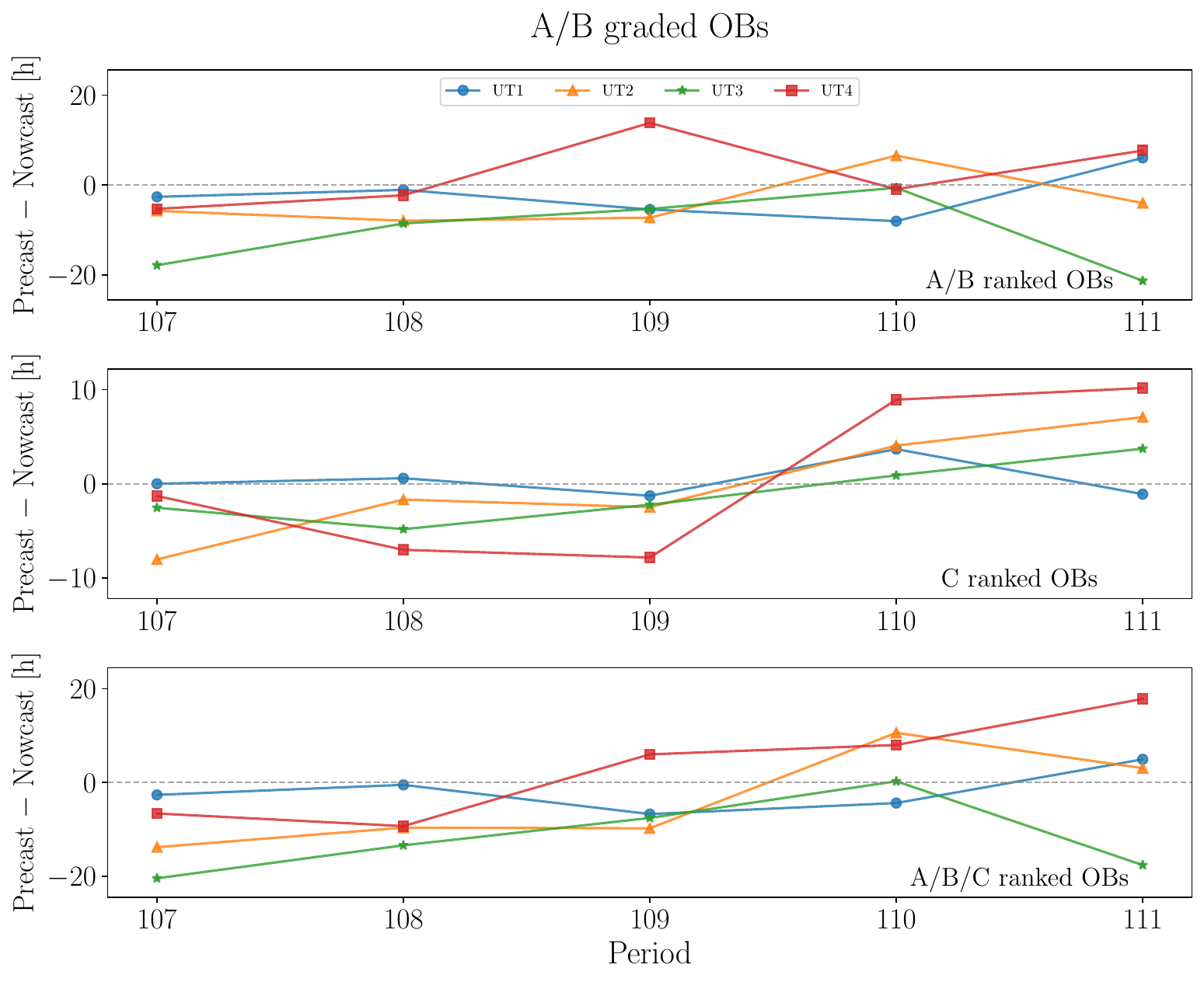}
\end{tabular}
\end{center}
\caption 
{\label{abab_1} 
Number of hours of A- or B-graded OBs executed in the simulation as a function of ESO observing semester, at each VLT UT, for precast minus nowcast. The top panel shows A- and B-ranked OBs, the middle panel C ranked, and the bottom panel the combined A/B/C-rank observations. Negative values indicate that the new model - nowcast - is performing better than the standard model, precast.}
\end{figure} 

\begin{figure} [ht]
\begin{center}
\begin{tabular}{c}
\includegraphics[height=11cm]{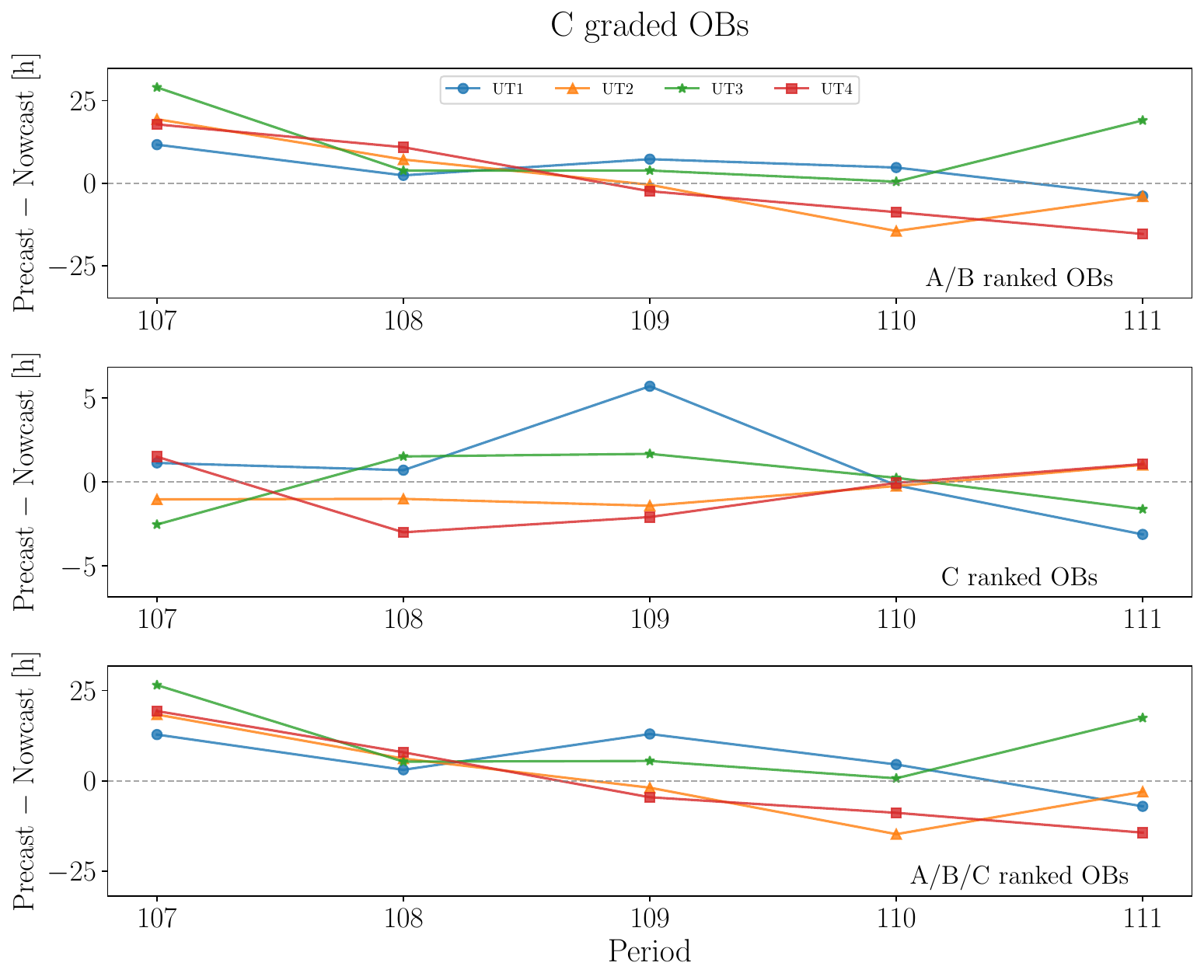}
\end{tabular}
\end{center}
\caption 
{\label{cab_1} 
Number of hours of C-graded OBs executed in the simulation as a function of ESO observing semester, at each VLT UT, for precast minus nowcast. The top panel shows A- and B-ranked OBs, the middle panel C ranked, and the bottom panel the combined A/B/C-rank observations. Positive values indicate that the new model - nowcast - is performing better than the standard model, precast.}
\end{figure}

In the first set of simulations presented in this section, we test one specific parameter – the method of providing the seeing input to OB filtering and ranking, while keeping all other parameters the same (OB queues from each semester, \href{https://www.eso.org/sci/facilities/paranal/quality-control/qc0-ob-grading.html}{standard QC0 policy}\footnote{Where an observations is considered a failure if the measured observational conditions are more than 10\%\ out of constraints.}, etc). That is, we test the different simulation results when using the precast (baseline seeing model) vs the nowcast (model to be tested). For the initial analysis presented, the results for all instruments at a specific UT are combined. We thus present the results per UT per semester over the range P107 to P111. The results are shown in Figure~\ref{abab_1} and Figure~\ref{cab_1}. 

Focusing on the A- and B-ranked OBs shown in the top panel of Figure~\ref{abab_1}, one can see that while there is clear dispersion between semesters and telescopes, the nowcast simulations appear to (on average) fall below the horizontal line equating to zero, meaning that the nowcast is – on average – more efficient at executing A- and B-ranked OBs. Then, Figure \ref{cab_1} shows (again focusing on the top panel of A/B ranked OBs) that the nowcast executed less C-graded OBs (falling above the horizontal). When summing over all semesters and telescopes the nowcast executes around 70 hours of more successful A/B-ranked OBs compared to precast, while executing around 88 hours less of OBs that are C-graded. When dividing by telescope and semester, this equates to around 3.5 hours per telescope, per semester of additional A/B-ranked successful OBs (with nowcast) and around 4.4 hours less C-graded OBs. 

Per telescope and per semester, this gain of around four hours of telescope time does not seem so significant. However, when starting this project and analysis and running these simulations, there was no guarantee that the nowcast could even do as well as the precast. The default, current PSO STS precast model was already a good model – it historically fails only around 10\% of the time. Thus, we believe that these results are very encouraging: our latest nowcast efforts are on par with – or are even outperforming – the precast model.  

As stated in Section~\ref{sec_metrics}, the metrics used above do not consider several important parameters that may affect our initial results and conclusions. The results presented in Figures~~\ref{abab_1} and~\ref{cab_1} compare the precast and nowcast success and failure in absolute hrs of OB executions. However, different simulations may run for different durations – dependent on how much time a simulation stays idle (when there is no observation to be executed under the current conditions). Such issues should be considered when more encompassing metrics are devised in the future.

\subsection{Nowcast vs Precast with constrained seeing}
\label{empiricalres}
\begin{figure} [ht]
\begin{center}
\begin{tabular}{c}
\includegraphics[height=11cm]{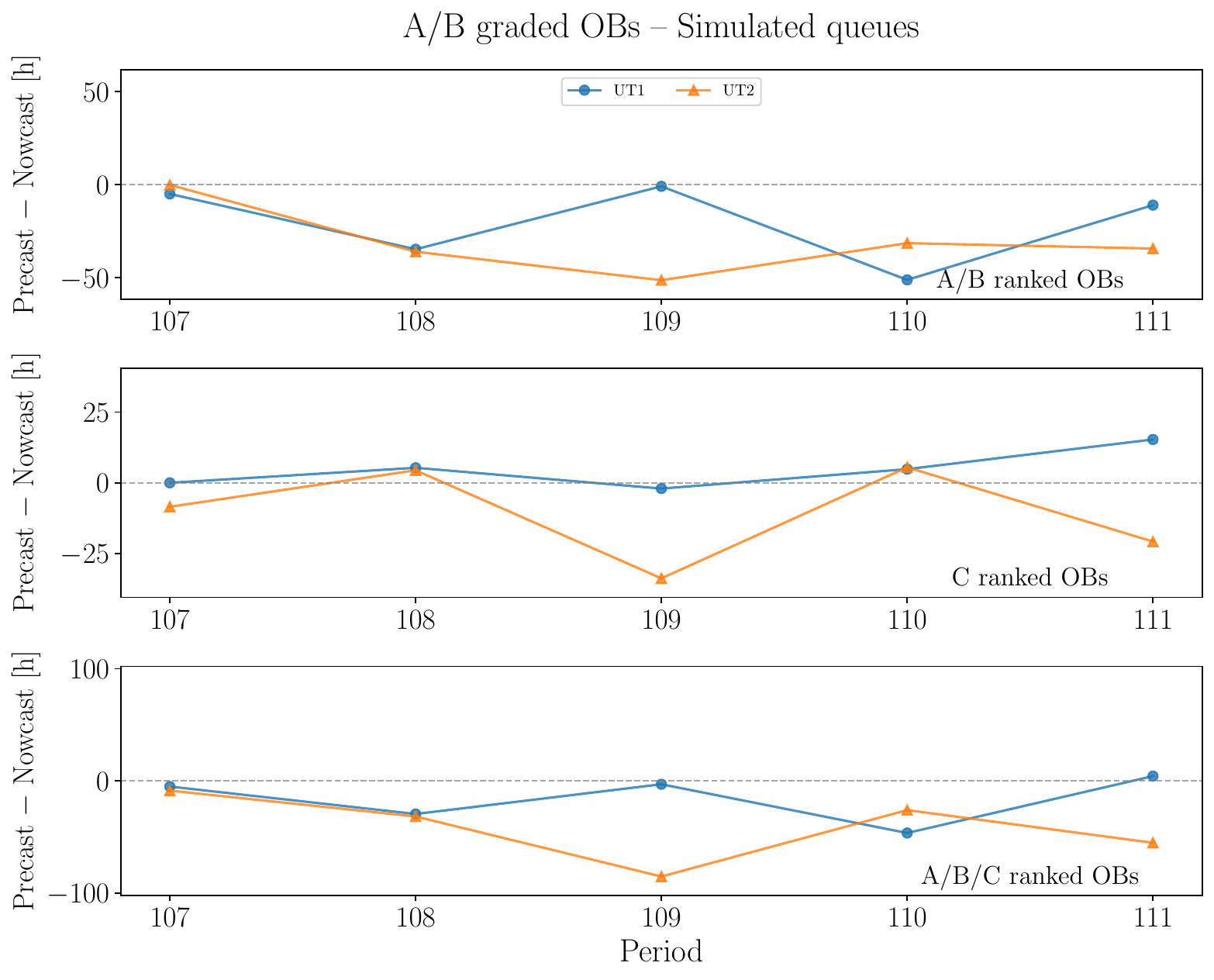}
\end{tabular}
\end{center}
\caption 
{\label{ut12_seeing_AB} 
Number of hours of A- or B-graded OBs executed in the simulation as a function of ESO observing semester, at UT1 and UT2, for precast minus nowcast, using OBs with IQ constraints that match the available seeing distribution. The top panel shows A- and B-ranked OBs, the middle panel C ranked, and the bottom panel the combined A/B/C-rank observations. Negative values indicate that the new model - nowcast - is performing better than the standard model, precast.}
\end{figure}

\begin{figure} [ht]
\begin{center}
\begin{tabular}{c}
\includegraphics[height=11cm]{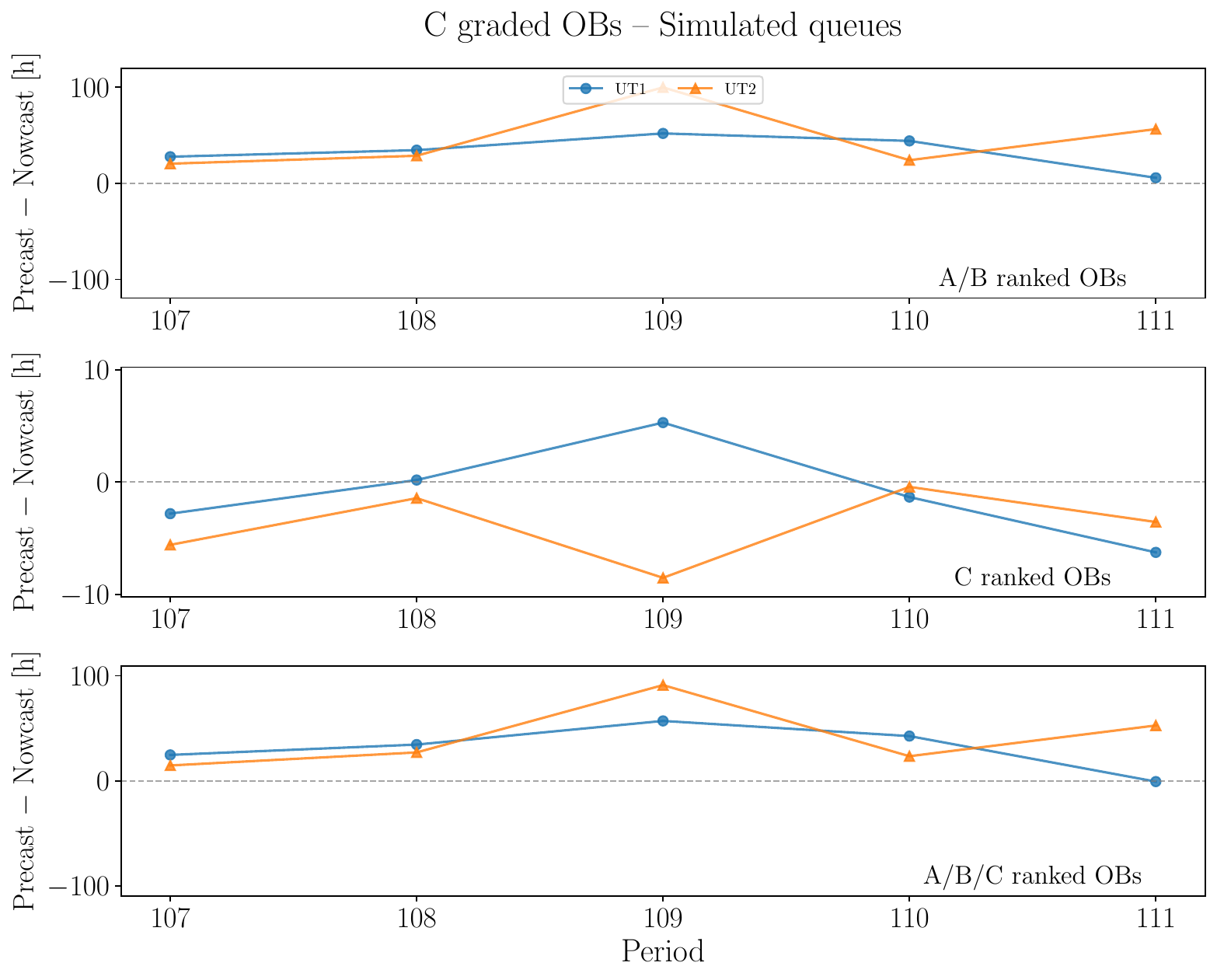}
\end{tabular}
\end{center}
\caption 
{\label{ut12_seeing_C} 
Number of hours of C-graded OBs executed in the simulation as a function of ESO observing semester, at UT1 and UT2, for precast minus nowcast, using OBs with IQ constraints that match the available seeing distribution. The top panel shows A- and B-ranked OBs, the middle panel C ranked, and the bottom panel the combined A/B/C-rank observations. Positive values indicate that the new model - nowcast - is performing better than the standard model, precast.}
\end{figure}

As shown in Figure~\ref{seeing}, the user-requested seeing (converted from requested IQ) distribution is significantly offset to larger values than the empirical, measured distribution at Paranal. Therefore, one may ask: how would the precast and nowcast models compare if users were setting more stringent constraints in their OBs? Such an analysis is also pertinent for future ELT observations, where one assumes that users will request more stringent observing conditions to exploit the unique resolving capabilities of adaptive optics on a 39m telescope. In this section we simulate such observations and assess differences in the results between precast and nowcast. To achieve this, the user-requested IQ constraint in each OB is replaced by a value randomly drawn from the empirical seeing distribution measured at Paranal. Then, we again run simulations using these new ‘mock OBs’. The IQ replacement and subsequent simulations are only done for UT1 and UT2. This was for simplicity: for UT3 and UT4 additional work is required to treat the coherence time and subsequent Turbulence Category (TC, combination of seeing and coherence time). Figures~\ref{ut12_seeing_AB} and~\ref{ut12_seeing_C} are the same as Figures~\ref{abab_1} and~\ref{cab_1}, but now using our mock OBs, with simulations restricted to UT1 and UT2 (and their respective instruments). Figure~\ref{ut12_seeing_AB} thus shows the number of successful OB hours executed for precast minus nowcast, while Figure~\ref{ut12_seeing_C} shows the hours of unsuccessful OBs for precast minus nowcast, both as a function of observing semester.

Differences between Figures~\ref{abab_1} and~\ref{cab_1}, and Figures~\ref{ut12_seeing_AB} and~\ref{ut12_seeing_C} are immediately obvious: the nowcast significantly outperforms precast both with respect to A/B-ranked successful OB execution (Figure~\ref{ut12_seeing_AB}, top panel) and C-graded A/B-rank OBs (Figure~\ref{ut12_seeing_C}, top panel). For more highly constrained OBs around 25 hours of additional A/B-ranked OBs are successfully executed each semester at each telescope (UT1 and UT2) when using nowcast (compared to precast), whereas nowcast leads to around 40 hours less C-graded OBs per telescope, per semester. This implies that the nowcast model is significantly more successful than the precast \textit{at executing more highly-constrained OBs, specifically those requesting lower seeing values}. Assuming that OBs submitted for ELT-instruments are indeed more constraining than the current distribution of VLT OBs, the use of nowcast in place of precast may indeed significantly increase STS efficiency for future integrated VLT+ELT operations.

\section{Conclusions and outlook}
\label{conclusions}
This contribution has provided a status report on simulations of PSO STS with specific emphasis on whether STS could be improved using a ML forecast of the seeing (nowcast) in place of the last 10\,min median (precast) as input to STS OB filtering. We have run two sets of simulations of OB executions between observing periods P107 and P111. When using ‘normal’ OBs (i.e. those with the user-provided IQ constraints) the nowcast slightly outperforms precast in terms of hours of successful and unsuccessful OB execution. While this result – in absolute hours of telescope time gained – is not so large, at a minimum it shows that we could implement a nowcast seeing model into STS today without any degradation in STS efficiency, and additionally we would probably gain several hours of telescope time at each telescope each semester. Before performing these simulations, it was not obvious that the nowcast would match the precast model – forecasting the weather/atmospheric parameters is notoriously difficult! That our current nowcast performs just as well as the precast, with suggestions that it can perform better, leads one to be optimistic about its use in future operations. This optimism is further supported by the work presented in the previous section. When one runs simulations using OBs with more stringent seeing values (those that match the available seeing at Paranal), nowcast significantly outperforms precast.  

Perhaps even more important than the above results with respect to whether we use precast or nowcast seeing input for STS, we now have a robust end-to-end simulator of the PSO STS operations model. The possible use cases of such a simulator go much beyond testing changes in the seeing-input model. We are now in the position to test all aspects of Paranal STS efficiency.
In the short term, we plan to assess how optimising for run completion (prioritising that observations from the same run/proposal are executed, once started) affects the overall STS efficiency. Naively, one may assume only positive outcomes of prioritising run completion. However, this may lead to other runs to never be started, or may lead to a prioritisation of runs with less stringent observing constraints. All of this can be tested with our STS simulator.

Our current STS model, specifically the use of accumulative probability distributions of the observing constraints, is currently applied under the assumption that user-supplied constraints populate the available observing conditions at Paranal. Figure~\ref{seeing} shows that this is clearly untrue for IQ/seeing, but this is also
the case for other paramaters such as observation coordinates, sky transparency request and Moon requests. This leads one to assume that optimisisng STS efficiency should consider the queue distributions of OB properties at any given time t when a ranking is done. This is currently being investigated. In addition, current PSO STS does not consider future scheduling constraints. On any given night, or time during that night, there exist future scheduling constraints that could be considered when deciding which observation to execute at time t. Such constraints include, e.g., there is a technical test scheduled for the next N nights, there is a visitor observing for the second half of the night, or - in more uncertain terms - there is a strong prediction of high winds (meaning the telescopes will be closed) for the next N hours. In principle one could include such variables to optimise STS. Again, these prospective STS changes can be simulated with the tools outlined in this paper, and a decision can be made on their possible implementation. 

For more than 20 years ESO has offered and executed an increasingly automated scheduling of observations during each night of available telescope time. In this contribution we have discussed the overall LPO STS model and speculated on how it could be further optimised. During the next years many different STS scenarios will be simulated, wth the aim of arriving to a model that successfully executes as many highly-scientifically-ranked observations as possible.

\bibliography{report} 
\bibliographystyle{spiebib} 

\end{document}